\documentclass[%
 reprint,
 amsmath,amssymb,
 aps,nofootinbib,
]{revtex4-1}
\usepackage[usenames, dvipsnames]{color}
\usepackage[autostyle]{csquotes}
\usepackage{pgfplots}
\usepackage{tikz}
\usepackage{graphicx}
\usepackage{dcolumn}
\usepackage{bm}
\usepackage{enumerate}

\begin{document}


\title{Inducing Gravity from Connections and Scalar Fields}
\author{Hemza Azri}
 \email{hm.azri@gmail.com}
\affiliation{%
Department of Physics, Ko\c{c} University\\ 34450 Sar{\i}yer, \.{I}stanbul, TURKEY}%


\begin{abstract}
We propose an approach to induced gravity, or Sakharov's \enquote{metrical elasticity}, which requires only an affine spacetime that accommodates scalar fields. The setup provides the induction of metric gravity from a \textit{pure affine} action, and it is established in two possible ways: (\textit{i}) at the classical level, Einstein-Hilbert action arises with both, metric and Newton's constant, from the nonzero potential energy of the background field (\textit{ii}) at the quantum level (quantized matter), gravity scale is induced from the one-loop effective action by integrating out the scalar degrees of freedom. In the former, the cosmological constant is absorbed leading to the gravitational sector, however, the fact remains that quantum corrections induce a large cosmological constant. This new approach adds a crucial feature to induced gravity which is the fact that the metric structure is not imposed from the scratch, but it is an outcome of the primary theory.  
\begin{description}
\item[Keywords]
Induced gravity; Affine gravity; One-loop effective action.
\end{description}
\end{abstract}

\pacs{Valid PACS appear here}
\maketitle


\section{Introduction}
\label{sec:1}
Despite the tremendous efforts that have been devoted to the quantum aspects of gravity, the shared conclusion has been simply that \textit{there is no viable description of quantum gravity} \cite{quantum-gravity}. On the other hand, electroweak and strong interactions of quarks and leptons are simply understood in the context of gauge field theory and are successfully encoded in the standard model (SM) of particle physics that has been completed thanks to the discovery of the Higgs boson \cite{LHC}. The gap existing between the physics of gravity and that of the SM has led some people to think about gravity as an emergent or an\enquote{induced} phenomenon that does not necessitate any quantum description, but rather, it emerges from manifestations of quantum matter fields \cite{sakharov,adler1,*adler2,*zee}. One of the attempts that gained much attentions has been the idea of induced gravity firstly proposed by Sakharov \cite{sakharov}. It is argued there that gravity as a classical force, could arise from quantum fields of elementary particles when treated in a \enquote{classical} curved spacetime. 

In induced gravity models, we generally suggest the existence of a geometric background; a \enquote{Lorentzian manifold} on which one proceeds to construct a quantum field theory. In this setup, the geometry is considered classical, whereas matter fields are quantized. This construction automatically generates Einstein's general relativity (GR) and higher curvature gravities.

The framework is generally based on the one-loop effective action of a classical theory of a massive scalar $\phi$ coupled nonminimally to the spacetime scalar curvature $R\left(g\right)$ as
\begin{eqnarray}
\label{metric-induced gravity action}
S[g,\phi]=\int d^{4}x \sqrt{-g}
\left[\frac{\xi \phi^{2}}{2}R(g)-\frac{1}{2}(\partial \phi)^{2}-V(\phi) \right]
\end{eqnarray}
where $g_{\mu\nu}$ refers to the metric tensor of the manifold and $\xi$ is a constant dimensionless parameter.
Historically the model above which is generically called an induced gravity model is a particular type of the so called scalar-tensor theory proposed originally by Jordan \cite{jordan} which stands on embedding the four-dimensional spacetime in a five-dimensional flat space and allows one to describe a variable gravitational constant \cite{maeda}. More influential are Brans-Dicke scalar-tensor gravities \cite{BD} which are also based on the same feature; the nonminimal coupling dynamics (see Ref.\cite{maeda} for both historical and technical details.) Among various and several examples, some earlier applications of scalar-tensor theories have included gravitational waves \cite{wagoner}, and later black hole structures \cite{bronnikov} as well as attempts to explain the accelerating expansion of the universe \cite{boisseau}.     
At the classical level, models such as (\ref{metric-induced gravity action}) have been used as an attempt to incorporate the spontaneous symmetry breaking in a curved background that leads to the scale of gravity \cite{adler1}. Furthermore, they have been considered as possible gravitational models for the early universe, particularly to drive cosmological inflation \cite{*early1,*early2,*early3,*early4}. 

The essential ingredient in Sakharov's induced gravity is the following one-loop effective action obtained from (\ref{metric-induced gravity action}) by integrating out the scalar degrees of freedom (the general case would include fermions and gauge bosons) \cite{sakharov,adler1}
\begin{eqnarray}
\label{metric one-loop}
\Delta S = \frac{i}{2}\text{Tr ln}\left[\Box_{g}+V^{\prime\prime} 
+\xi R(g)
\right].
\end{eqnarray}

Although it leads correctly to Einstein's general relativity and generalized theories of gravity, this ancient induced gravity may not reflect a correct emergence of gravity based on its \enquote{metrical} structure. In fact, classical gravity described by Einstein's general relativity is a theory of the spacetime metric. The latter plays the role of the gravitational field which in turn appears as curvature effects on rods and clocks. These effects are encoded in the metric tensor, and then it is this \enquote{metrical elasticity} which is the origin of gravity at large scales. At that end, a correct theory of induced gravity must be able to generate this metrical elasticity of space. In the standard induced gravity, this metric structure is already postulated as a Lorentzian manifold, thus generation of Einstein-Hilbert action may not mean generation of \enquote{metrical elasticity}.

The question now, is it possible to generate theories of gravity with their metrical structure? If yes, do they arise \enquote{classically} or at the quantum level like standard induced gravity where matter is quantized?

The aim of this paper is to show that such a theory might be possible in the framework of \enquote{affine} spacetime, where the metric tensor is completely absent in the beginning \cite{azri-induced}. Our setup will be based on an affine action in which an affine connection, introduced as fundamental quantity, is coupled to scalar fields through curvature. We will propose to proceed in two possible ways:
\begin{enumerate}[A.]
\item The first step that we follow in Section \ref{sec:2} is to expand our pure affine action around a background up to quadratic terms of the field. Stationary of the action against variation with respect to the affine connection leads to dynamical equation from which emerges the metric tensor with a constant of proportionality that will be related to the gravitational constant. As a result, the action of the background field is reduced to Einstein-Hilbert action where any constant piece that appears \textit{a priori} in the potential, which correspond to a possible cosmological constant, is completely absorbed in the definition of this gravitational sector. On the other hand, the terms quadratic in the field will bring higher order curvature terms that correspond to extended gravity models. This terminates the classical part of our approach to induced gravity. 
\item
The second part will be devoted to constructing the one-loop effective action of the theory. This maybe considered as the first essay towards effective actions in affine space. It is the emergence of the metric structure in (A) that allows us to calculate this quantity. In fact, the volume elements in affine space which are necessary in performing the integrals are reduced to metric volume elements thanks to step (A). We then adopt an ultraviolet (UV) cutoff to regularize the integrals which in turn lead to an induced gravitational and cosmological constants. As in Sakharov's induced gravity, the experimental value of Newton's constant requires a UV boundary of the order of the Planck mass. This, unfortunately, pushes the induced cosmological constant to a very large value due to the presence of the quartic and quadratic UV divergent terms.  
\end{enumerate}
With the last step, we have a complete approach to induced gravity where not only the gravitational parameters (Newton and cosmological constants) but interestingly, the metric tensor also gains an emergent character.
A summary of these results will be given in Section \ref{sec:3}.

\section{Generating metric gravity}
\label{sec:2}
As we have mentioned above, our aim is to construct a model of induced gravity with a crucial feature; the induction of the spacetime metric. This suggests a primary \enquote{metric-less} classical action from which one proceed to extract the one-loop effective action. Thus, we begin with a very simple spacetime which does not recognize a metrical structure where angles and distances measurements take place. However, different events in different points are studied and compared only through parallel displacements of vectors and tensors \cite{azri-induced,azri-affine,demir-eddington,*azri-immersed,*azri-separate,*poplawski,*cota,*oscar,*oscar2}. 
To that end, this spacetime arena is endowed with only an affine connection $\Gamma$ and its associated curvature. This simple structure offers a viable calculus of variation by coupling scalar fields to the affine connections \cite{kijowski1,*kijowski2}. Indeed, a viable theory of gravity must be described by a covariant field equations that arise from an action principle.
Here we want to construct an affine action that enables us to induce the scale of gravity in the philosophy of GR counterpart (\ref{metric-induced gravity action}), i.e, when $ \xi \phi^{2}$ $\rightarrow $ $M_{Pl}^{2} $ (for constant fields.) With this, the general coordinate transformations of affine spacetime suggest then a combination of the Ricci tensor of the affine connection $R_{\mu\nu}(\Gamma)$ and scalar field kinetic structure $\nabla_{\mu}\phi\nabla_{\nu}\phi$ as well as a scalar potential $V(\phi)$.
Dimensional analysis implies that this combination could simply come out in the following form 
\begin{eqnarray}
\label{action0}
S \left[\phi \right]= 2\int d^{4}x \frac{\sqrt{\left| \right| \xi \phi^2 R_{\mu\nu}\left(\Gamma\right)- \nabla_{\mu}\phi\nabla_{\nu}\phi \left| \right|}}{V(\phi)},
\end{eqnarray}
which indeed leads to GR with an induced gravity scale for constant fields \cite{azri-induced} (see also Ref.\cite{azri-review} on how to construct pure affine actions.)

We notice here that we take only the symmetric part of the Ricci Tensor of the symmetric affine connection.

This action has been introduced for the first time by the author and his collaborator as an attempt to a new approach to induced gravity in the philosophy of spontaneous symmetry breaking \cite{azri-induced}. It is worth enlightening briefly the main results of the mentioned work. First of all, this action runs along two important and explicit properties:  
\begin{enumerate}[(a)]
\item
Both geometry and scalar field terms define the invariant volume measure, i.e, the square root of the determinant. The scalar field enters this measure by its derivative (kinetic part) in a tensorial form. The property of \enquote{contraction} is absent at this stage since there is no notion of metric tensor.
\item 
The potential energy enters the action separately in division, and then the action is well defined only for nonzero potential energy, $V\left(\phi \right)\neq 0$.
\end{enumerate}

The second property reflects the viability of the affine models in studying the early universe where the scalar field $\phi$ requires a nonzero potential enegry to get all the phase of inflation done \cite{azri-induced,azri-affine,guth,*linde1,*albrecht,*linde2,*higgs-inflation,*bauer1,*bauer2}. The crucial importance of this property will become clear later when deriving the gravitational actions based on the model given in (\ref{action0}). It has been shown that (affine) gravity is induced via spontaneous symmetry breaking where the gravity scale arises from the constant vacuum expectation value of a heavy scalar, and the metric tensor is generated thanks to the nonzero vacuum energy left after symmetry breaking \citep{azri-induced}.    
 
Returning to the present work, the first step towards the emergence of metric gravity is to expand the field $\phi$ around a \enquote{constant} background $\phi_{c}$ as
\begin{eqnarray}
\label{field expansion}
\phi=\phi_{c}+\varphi.
\end{eqnarray}
This has been taken only for the sake of simplicity and it could be trivially generalized to the case where the background is not constant.
 
An important remark here, is that the field must not have a zero potential at this background, $V\left(\phi_{c}\right) \neq 0$. This is an essential postulate for the upcoming results.

To that end, and up to second order in $\varphi$, the action (\ref{action0}) takes the following form
\begin{eqnarray}
\label{expanded action}
S\left[\varphi+\phi_{c}\right]=
S\left[\phi_{c}\right]+I_{1}[\varphi]+I_{2}[\varphi\,\partial^{2}\varphi]+I_{3}[\varphi^{2}]+\dots \nonumber \\
\end{eqnarray}
where the first term in the right hand side is nothing but action (\ref{action0}) evaluated at the background $\phi_{c}$, and the last three terms are respectively given as follows 
\begin{widetext}
\begin{eqnarray}
\label{i1}
&&I_{1}[\varphi]=
2\int d^{4}x 
\frac{\sqrt{||K(\phi_{c})||}}{V^{2}(\phi_{c})}\Big[\xi \phi_{c} V(\phi_{c})
(K^{-1})^{\alpha\beta}R_{\alpha\beta}-V^{\prime}(\phi_{c})\Big]\varphi
\\&&
I_{2}[\varphi\,\partial^{2}\varphi]=\int d^{4}x\,\varphi\, \partial_{\alpha}
\left( \frac{\sqrt{||K(\phi_{c})||}}{V(\phi_{c})} (K^{-1})^{\alpha\beta}\partial_{\beta}\varphi
\right),
\label{i2}
\end{eqnarray}
and
\begin{eqnarray}
\label{i3}
I_{3}[\varphi^{2}]&&=
\int d^{4}x 
\frac{\sqrt{||K(\phi_{c})||}}{V^{2}(\phi_{c})}
\left[ \left(\xi V(\phi_{c})-2\xi \phi_{c}V^{\prime}(\phi_{c}) \right)
(K^{-1})^{\alpha\beta}R_{\alpha\beta}-2\xi\phi_{c}^{2}V(\phi_{c})
(K^{-1})^{\alpha\beta}
(K^{-1})^{\nu\lambda}R_{\beta\nu}R_{\lambda\alpha}
 \right] \varphi^{2} \nonumber \\&&
+\int d^{4}x 
\frac{\sqrt{||K(\phi_{c})||}}{V^{2}(\phi_{c})} 
\left[\frac{2V^{\prime 2}(\phi_{c})}{V(\phi_{c})}-V^{\prime \prime}(\phi_{c})
+\xi^{2}\phi_{c}^{2}V(\phi_{c})\left((K^{-1})^{\alpha\beta}R_{\alpha\beta}\right)^{2} \right]
\varphi^{2}
\end{eqnarray}
\end{widetext}
where for simplicity we have introduced the tensor $K_{\mu\nu}$ which is given as follows
\begin{eqnarray}
\label{tensor k}
K_{\mu\nu}\left(\phi\right)=\xi \phi^2 R_{\mu\nu}\left(\Gamma\right)- \nabla_{\mu}\phi\nabla_{\nu}\phi.
\end{eqnarray}
Finally we have an expansion of action (\ref{action0}) up to quadratic terms of the scalar field where all the geometric parts (the Ricci tensor and its inverse) that appear in this expansion are given only in terms of the affine connection.

Our next step is to extract the new feature behind this expansion which will be certainly the \textit{induction} of metric gravity (GR and its extensions). In the subsequent section we show how Einstein-Hilbert action arises after the \textit{generation} of the metric tensor dynamically. This classical setup will be of a great importance since it helps, later on, in performing the one-loop effective action which is considered as the main part in inducing gravity \`a la Sakharov.  
 
\subsection{Classical setup: Inducing metric structure}
First of all, let us simplify the expression of the expansion (\ref{expanded action}) by using every possible equation of motion. 

Here, the term (\ref{i1}) which is linear in $\varphi$ vanishes by use of the equation of motion of the background
\begin{eqnarray}
\frac{\delta S[\phi_{c}]}{\delta \phi_{c}}=0,
\end{eqnarray}
which is given explicitly as
\begin{eqnarray}
\label{equation of motion of phi}
\frac{\sqrt{||K(\phi_{c})||}}{V^{2}(\phi_{c})}\left[\xi \phi_{c} V(\phi_{c})
(K^{-1})^{\alpha\beta}R_{\alpha\beta}-V^{\prime}(\phi_{c})\right]=0
\end{eqnarray}
This equation of motion can be simplified using the geometrical equations which arises from varying the background action with respect to the affine connection $\Gamma$. In fact, stationary of $S[\phi_{c}]$ against this variation 
\begin{eqnarray}
\frac{\delta S[\phi_{c}]}{\delta \Gamma^{\lambda}_{\,\mu\nu}}=0,
\end{eqnarray}
would simply lead to the following constraint
\begin{eqnarray}
\label{dynamical eq}
\nabla_{\mu}\left(\xi \phi_{c}^{2} \frac{\sqrt{\left| \right| K\left(\phi_{c}\right) \left| \right|}}{V(\phi_{c})}
\left( K^{-1}\right)^{\alpha\beta} \right)=0.
\end{eqnarray}
This implies a natural \enquote{induction} of an invertible (rank-two) tensor $g_{\mu\nu}$ such that
\begin{eqnarray}
\label{metric}
\sqrt{\left| \right| g \left| \right|}\left(g^{-1} \right)^{\alpha\beta}=\frac{\xi \phi_{c}^{2}}{M^{2}} \frac{\sqrt{\left| \right| K\left(\phi_{c}\right) \left| \right|}}{V(\phi_{c})}
\left( K^{-1}\right)^{\alpha\beta},
\end{eqnarray}
where $M$ is a constant of a mass dimension.

The induced tensor clearly satisfies $\nabla_{\mu} g_{\alpha\beta}=0$ which appears as a metricity equation, and then it plays the role of the metric tensor. The affine connection $\Gamma$ in turn, which defines the action (\ref{action0}) is automatically reduced to Levi-Civita of this metric. 

With this induced metrical structure described by the metric tensor (\ref{metric}), the equation of motion of the background (\ref{equation of motion of phi}) takes the form
\begin{eqnarray}
\xi \phi_{c} R(g)-V^{\prime}(\phi_{c})+\Psi(\phi_{c})=0,
\end{eqnarray}
where $R(g)$ is now the Ricci scalar of the metric tensor and the last term is given by
\begin{eqnarray}
\Psi(\phi_{c})=\left(1- \frac{M^{2}}{\xi\phi_{c}^{2}}\right)V^{\prime}(\phi_{c}).
\end{eqnarray}
We mention again that the background field has been taken constant, the case that justifies the absence of the d'Alembert operator in the previous equation \citep{azri-induced}. At the background, the equality (\ref{metric}) is nothing but the gravitational field equations with a cosmological constant $V_{0}=V(\phi_{c})$ or explicitly
\begin{eqnarray}
\label{einstein equation with cc}
\xi\phi_{c}^{2}R_{\mu\nu}(g)=g_{\mu\nu}V_{0}\left(\frac{M^{2}}{\xi\phi_{c}^{2}} \right).
\end{eqnarray}
Consistency with Einstein's field equations implies that
\begin{eqnarray}
M^{2}=\xi\phi_{c}^{2}\equiv (8\pi G_{N})^{-1}.
\end{eqnarray}
Now the remaining terms in the expansion (\ref{i2}), (\ref{i3}) as well as $S\left[\phi_{c}\right]$ are reduced to quantities that depend on the induced metric and they lead finally to the gravitational actions  
\begin{eqnarray}
\label{induced action}
S\left[\varphi+\phi_{c}\right]&&= \int d^{4}x \sqrt{\left| \right| g \left| \right|}\,
\frac{R\left(g\right)}{16\pi G_{N}}
\nonumber \\ &&
+\int d^{4}x 
\sqrt{\left| \right| g \left| \right|}\, \, \varphi\, \Big[\Box_{g} +c_{1}+c_{2} R\left(g\right)\Big]\varphi \nonumber \\ &&
+ \int d^{4}x 
\sqrt{\left| \right| g \left| \right|}\, \, \varphi^{2}\, \Big[
c_{3} R^{2}\left(g\right)+
c_{4} R^{\mu\nu}R_{\mu\nu} \Big] \nonumber \\
\end{eqnarray}
where the constants $c_{i}$ are listed as follows
\begin{eqnarray}
\label{c1}
&&c_{1}=\frac{2V^{\prime 2}(\phi_{c})}{V(\phi_{c})}-V^{\prime \prime}(\phi_{c})
\\
&&c_{2}=\xi - \frac{2\xi \phi_{c} V^{\prime}(\phi_{c})}{V(\phi_{c})} \\
&&c_{3}=\frac{\xi\phi^{2}_{c}}{V(\phi_{c})} \\
&&c_{4}=-\frac{2\xi\phi^{2}_{c}}{V(\phi_{c})}.
\end{eqnarray}
It is worth noticing that in order to get the Ricci scalar in the first term of (\ref{induced action}), we have used equation (\ref{metric}) in its tensor form where the potential $V(\phi_{c})$ which appears in $S[\phi_{c}]$ is then written in terms of the Ricci scalar\footnote{Here, the explicit relation between $V_{0}$ and $R(g)$ is obtained by tracing equation (\ref{einstein equation with cc}). The author thanks D. Demir for drawing his attention to this step.}. 

The induced actions (\ref{induced action}) are written in terms of the metric tensor and then they describe metrical gravitational theories. These metrical theories are originated from the simple and primarily action (\ref{action0}). We thus enumerate the main results found above as follows:
\begin{enumerate}[(i)]
\item Einstein-Hilbert action which describes GR is \textit{generated} from the (constant) background field via the action $S\left[\phi_{c}\right]$. This is nothing but Eddington (affine) action written in an associated metrical form \cite{eddington,*schrodinger}. A nonzero potential is a very crucial condition here. Here, we notice the absence of any constant term that could correspond to a \enquote{classical} cosmological constant. The reason is that at the background field, the constant potential $V_{0}=V(\phi_{c})$ is \enquote{completely} transformed to the gravitational sector which is formed by the Planck mass and the Ricci scalar. Thus, the affine to GR transition is followed by the absorption of the cosmological constant.     
\item
This pure affine model (\ref{action0}) generates also higher order curvature terms which are seen in the last integral of (\ref{induced action}). Unlike metric induced gravity, these terms appear here in the action without calculating the effective action. In metric induced gravity the higher order curvature terms arise only after adopting an explicit UV cutoff and regularizing the action. The one-loop effective action will be our object of interest in the subsequent Section. 
\end{enumerate}
\subsection{Quantum corrections: Gravity \`a la Sakharov}

Up to now, like geometry, we have treated the matter fields (scalars here) as pure classical fields. In what follows, we will define the quantum matter fields on the classical background (affine) geometry. The aim of this part of work is to consider the effects of quantum matter fields on this geometry which is endowed with only an affine connection and its curvature.

The object of interest here is the effective action $\Delta S[\phi]$ for (\ref{action0}) by integrating out the scalar fields. The one-loop contribution to this effective action is generally given by \cite{peskin,*davies,*buch}
\begin{eqnarray}
\label{affine one-loop}
e^{i\Delta S[\phi]} \equiv 
\int \mathcal{D}\varphi \exp \Big[ \frac{i}{2}\varphi \left(\frac{\delta^{2}S[\phi]}{\delta\phi\delta\phi} \right)\varphi \Big].
\end{eqnarray}
As one may easily show, the integrand in this expression is given by the sum of the two terms (\ref{i2}) and (\ref{i3}) quadratic in the field. Thus, the expression (\ref{affine one-loop}) is a function of the affine connection and its associated curvature and it can be considered as the effective action for the classical \textit{affine connection} $\Gamma_{\,\mu\nu}^{\lambda}$ which appears (in terms of Feynman diagrams) in the external legs, whereas the scalar fields inside the loops. However, it is worth emphasising here that, as we shall see below, this method will be based on a transition from pure-affine to pure-metric action where the metric tensor gains an induced character from the background action. Thus, one has to pay attention to that and not confuse this case with the general classical cases in Ref.\cite{azri-affine,azri-induced} where the metric tensor does not arise from ''only'' the background but satisfies a complete Einstein equations of motion with scalar fields\footnote{By this we mean that the present method of induced gravity has nothing to do with any quantum corrections of the pure-affine inflationary models given in Ref.\cite{azri-affine,azri-induced,azri-review}.}.  

As we have mentioned in the previous part of this section, the geometric side of the equations of motion, or the equations arising from variation with respect to the affine connection leads to dynamical constraint (\ref{dynamical eq}) which in turn necessitates the existence of an emergent metric field (\ref{metric}). This step adds a new feature to induced gravity in general which is the classical transition
\begin{center}
\textit{Affine spacetime} $\xrightarrow{\text{Eq}.(\ref{dynamical eq})}$ \textit{Metric spacetime}.
\end{center} 
The important utility of this transition in evaluating the effective action, is the fact that spacetime affine measure is reduced to spacetime metric measure, or  
\begin{eqnarray}
\frac{\xi^{2}\phi_{c}^{4}}{M^{4}}\frac{\sqrt{||K(\phi_{c})||}}{V^{2}(\phi_{c})}
\xrightarrow{\text{Eq}.(\ref{metric})} \sqrt{||g||}.
\end{eqnarray}
To that end, in our approach, gravity will be induced from the one-loop effective action (\ref{affine one-loop}) which takes the form
\begin{eqnarray}
e^{i\Delta S[\phi]} \rightarrow
\int \mathcal{D}\varphi \exp \Big\lbrace  i
\int d^{4}x \sqrt{||g||}\,
\varphi \Big[\Box_{g} +\mathcal{H}(g) \Big]\varphi \Big\rbrace \nonumber \\
\end{eqnarray}
where the operator $\mathcal{H}(g)$ is given by
\begin{eqnarray}
\mathcal{H}(g)=c_{1}+c_{2} R\left(g\right) 
+c_{3}R^{2}\left(g\right)+c_{4}R^{\mu\nu}R_{\mu\nu}.
\end{eqnarray}
This simply leads to\footnote{The final result includes a factor two in front of the operator $\Box_{g} +\mathcal{H}(g)$ in order to have a correct Gaussian integral. However, we have ignored this factor since the quantity of interest will be difference
in the one-loop contribution to the effective actions of two metrics using the familiar formula of $\ln(a/b)$ (see References \cite{visser,extended-sakharov} for details.)}
\begin{eqnarray}
\label{final one-loop}
\Delta S \rightarrow \frac{i}{2}\,\text{Tr}\ln \left[\Box_{g} +\mathcal{H}(g)\right]. 
\end{eqnarray}
This is the final expression of the one-loop effective action that arises from pure affine action (\ref{action0}). There are crucial differences from the one-loop effective action (\ref{metric one-loop}) in metric theory which could be seen if one considers a simple massive scalar field (see discussion below).

A common fact however is that those expressions diverge when performing the integration. The detailed analysis behind the evaluation of the integrands (\ref{metric one-loop}) and (\ref{final one-loop}) put it beyond the scope of this paper. However, we mention here that we follow the Heat Kernel method where the heat kernel is expressed in terms of the Seeley-DeWitt expansion, and the integrals are regulared by adopting a UV cutoff \cite{davies,buch}. 

Below, we will not be interested in the higher order (curvature) terms but only in the divergences which are proportional to the Ricci scalar $R$ and that corresponding to a volume element. Detailed calculation shows that those two terms \textit{induce} both Newton's constant $G_{\text{ind}}$ and a cosmological constant $V_{0}^{\text{ind}}$ respectively  
\begin{eqnarray}
\label{induced-G}
\frac{1}{G_{\text{ind}}}=&&\left(\frac{1}{6}-c_{2} \right)\frac{\Lambda_{\text{UV}}^{2}}{2\pi} \nonumber \\
&&-\left(\frac{1}{6}-c_{2} \right)\frac{c_{1}}{\pi}
\ln\left(\frac{\Lambda_{\text{UV}}}{\mu} \right)\nonumber \\
&&+\text{UV-finite}
\end{eqnarray}
and
\begin{eqnarray}
\label{induced-cc}
V_{0}^{\text{ind}}=&&
-\frac{\Lambda_{\text{UV}}^{4}}{128\pi^{2}} \nonumber \\
&&+c_{1}\frac{\Lambda^{2}_{\text{UV}}}{64\pi^{2}}-\frac{c_{1}^{2}}{64\pi^{2}}
\ln\left(\frac{\Lambda_{\text{UV}}}{\mu} \right)\nonumber \\
&&+\text{UV-finite},
\end{eqnarray}
where the parameter $\mu$ is an infrared cutoff.

These expressions show the main idea behind induced gravity which lies in determining the gravitational couplings from the particle spectrum and the UV cutoff. The latter, however, will be fixed by the numerical values provided by experiments and it will certainly be related to the upper scale for the valid effective theory. For the previous setup to make physical sense, the UV boundary must be of the order of the Planck mass $\Lambda_{\text{UV}} \simeq M_{Pl}$. Thus, the induced gravitational constant reads 
\begin{eqnarray}
\frac{1}{G_{\text{ind}}} \simeq M_{Pl}^{2}.
\end{eqnarray}
This has a direct but sever implication on the value of the cosmological constant. While its observational value is estimated to be of the order of the neutrino mass density, i.e, $(10^{-3} \text{eV})^{4}$ \cite{planck}, the value of the induced quantity (\ref{induced-cc}) manifests as 
\begin{eqnarray}
|V_{0}^{\text{ind}}| \simeq M_{Pl}^{4} \simeq (10^{19}\text{GeV})^{4}.
\end{eqnarray}
The discrepancy between the observational and theoretical estimations of the value of the cosmological constant is not restricted to the present model of induced gravity but it holds in most of gravity models albeit with quantum matter fields. This is simply the origin of the celebrated cosmological constant problem \cite{ccp1,*ccp2}. 
\begin{table*}[t]
\caption{\label{table1}%
The one-loop induced parameters in standard (metric) induced gravity versus those of the present (affine) approach for quadratic potential. In both approaches, the gravitational and the cosmological constants are dominated by power-law UV terms, however, in those terms the conformal value $\xi=1/6$ of metric gravity is no longer preserved in our approach. This could be a consequence of the absence of (metric) conformal transformation in affine gravity \cite{azri-review}. The crucial differences between the approaches lie in the induced higher order gravity where the couplings are power-law UV sensitive in the present model. This is an expected result since higher order gravity emerges classically in (\ref{induced action}) before performing the effective action. }
\begin{ruledtabular}
\begin{tabular}{cccc}
\textrm{Induced constants}&
\textrm{Metric induced gravity}&
\textrm{Affine induced gravity}\\
\colrule
\\
$V_{0}^{\text{ind}}$ & $-\frac{1}{64\pi^{2}}\left[\frac{\Lambda^{4}}{2}-m^{2}\Lambda^{2}-m^{4}\ln(\Lambda/\mu) \right]$ & $-\frac{1}{64\pi^{2}}[\frac{\Lambda^{4}}{2}-3m^{2}\Lambda^{2}+9m^{4}\ln(\Lambda/\mu) ]$\\ 
\,&\,&\, \\
$1/G_{\text{ind}}$ & $\frac{1}{\pi}(\frac{1}{6}-\xi)\left[\frac{\Lambda^{2}}{2}-m^{2}\ln(\Lambda/\mu)\right]$ & $\frac{1}{\pi}(\frac{1}{6}+3\xi)\left[\frac{\Lambda^{2}}{2}-3m^{2}\ln(\Lambda/\mu)\right]$\\
\,&\,&\, \\
$c_{_{R}}$ & $\frac{1}{32\pi^{2}}\left(\frac{1}{36}-\frac{\xi}{3}+\xi^{2}\right)\ln(\Lambda/\mu)$ & $\frac{1}{32\pi^{2}}\left[-2\xi^{2}\frac{\Lambda^{2}}{m^{2}}+ \left(\frac{1}{36}+13\xi+9\xi^{2}\right)\ln(\Lambda/\mu)\right]$  \\
\,&\,&\, \\
$\bar{c}_{_{R}}$ &$-\frac{1}{2880\pi^{2}}\ln(\Lambda/\mu)$ & $\frac{1}{32\pi^{2}}\left[4\xi\frac{\Lambda^{2}}{m^{2}}- \left(\frac{1}{90}+24\xi\right)\ln(\Lambda/\mu)\right]$ \\
\end{tabular}
\end{ruledtabular}
\end{table*}
Last but not least, we summarize in Table \ref{table1} the UV coefficients of the induced gravitational and cosmological constants, as well as the higher order curvature terms, for the simple example of a massive scalar field where $V(\phi)=m^{2}\phi^{2}/2$ using the following standard writings
\begin{eqnarray}
\label{example}
\Delta S \supset
\int d^{4}x \sqrt{||g||}\,  \Big[\frac{R(g)}{16\pi G_{\text{ind}}}
-2V_{0}^{\text{ind}}
+c_{_{R}}R^{2} \nonumber \\
+\bar{c}_{_{R}}R_{\mu\nu}R^{\mu\nu}+\dots
\Big].
\end{eqnarray}
Here, since curvature terms like those at the end of the last expression are also generated classically in (\ref{induced action}), then the related coefficients will be certainly sensitive to UV power-law. This is no longer the case of metric induced gravity.
 
The discussion above shows how one could realize an induced gravity model from a pure affine spacetime free of not only the gravitational constant which provides the measure of weights, but also the metric tensor which plays the role of the gravitational field.  The present setup is able to induce both quantities as well as a vacuum energy from the one-loop (affine) effective action. 

We conclude by mentioning that the present approach may need to be extended by illuminating the following points:
\begin{enumerate}[i.]
\item First of all, the model described by action (\ref{action0}) is based mainly on scalar degrees of freedom and it could be easily generalized to multi-scalar fields \cite{azri-review, azri-thesis}. However, the fact remains that fermionic fields are not yet viably accommodated in affine space. This could be an obstacle if the gravitational couplings have to be induced from \enquote{all} the SM particles spectrum. In this case, extending (\ref{action0}) to include the SM matter fields in a unified picture would be necessary.
\item Second, unlike the case of induced gravity models based on their prior metric structure, in the present approach, we were able to generate higher order gravity terms only classically. Although these terms receive a UV power laws corrections (see Table \ref{table1}), one has to explore the possible and considerable new features of these terms, such as deriving cosmological inflation.        
\end{enumerate}   

\section{Summary}
\label{sec:3}

The physics of gravity at the microscopic scale remains one of the puzzles on which debates have never been settled down. Since the standard model of particle physics does not accommodate this interaction, people turned to thinking about its origin as a non-fundamental force. Gravity is induced or emerged from the micro-physical phenomena is the common framework for most of the proposed models \cite{sakharov,extended-sakharov1}.

Induced gravity, particularly in Sakharov's approach, aims at describing classical gravity by the quantum fields of matter. The latter need to be coupled to spacetime curvature with no signature of Newton's constant that translates the presence of gravity. At this stage, the \enquote{classical} spacetime geometry does not show to have any specific dynamics. Nevertheless, its response to quantum matter fields ends up with an induction of the gravitational constant via the one-loop effective action. In Sakharov's approach, the spacetime gains a Riemannian geometry from the scratch, while in other (almost) similar models, it has been taken arbitrary \cite{extended-sakharov}. However, in all these models, the main postulate is that spacetime is \enquote{endowed} with a metric tensor.    

In this paper we have argued that gravity, if induced, must arise with the metrical structure of spacetime. The latter property may not be imposed from the scratch. In fact, it has been shown that matter fields may not require the metric tensor to couple to curvature of spacetime \cite{azri-induced,azri-affine,kijowski1,kijowski2}. It turned out that the metric tensor is generated dynamically from an affine variational principles. This feature has been at the heart of our approach presented in this paper where we have proposed a pure affine action in which a scalar field is coupled to the affine connection. Unlike the models that we have referred to previously, our approach is founded on two main roads, where in the first, we have generated the metric structure dynamically. Then we have proceeded to calculating the one-loop effective action and obtained the regularized induced parameters; Newton's constant and the cosmological constant, providing that the ultraviolet cutoff is comparable to the Planck mass.

Induced gravity does not only provide us with an origin to the gravitational couplings from the particle spectrum, but it may also have an interesting effects on the standard model of particle physics particularly in counteracting the ultraviolet sensitivity of the latter \cite{demir-uv0,*demir-uv1,*demir-uv2,*demir-uv3,*demir-uv4}. The present approach to induced gravity may reveal some important and new features of the physics of standard model when incorporating gravity. This will be explored in a possible future work.      

\section*{acknowledgments}
The author thanks Durmu\c{s} Demir and Tomi Koivisto for their constructive comments. Supports (in part) by T\"{U}B\.{I}TAK grant 117F111 are acknowledged.

\bibliographystyle{apsrev4-1.bst}
\bibliography{ref}

\begin{thebibliography}{52}%
\makeatletter
\providecommand \@ifxundefined [1]{%
 \@ifx{#1\undefined}
}%
\providecommand \@ifnum [1]{%
 \ifnum #1\expandafter \@firstoftwo
 \else \expandafter \@secondoftwo
 \fi
}%
\providecommand \@ifx [1]{%
 \ifx #1\expandafter \@firstoftwo
 \else \expandafter \@secondoftwo
 \fi
}%
\providecommand \natexlab [1]{#1}%
\providecommand \enquote  [1]{``#1''}%
\providecommand \bibnamefont  [1]{#1}%
\providecommand \bibfnamefont [1]{#1}%
\providecommand \citenamefont [1]{#1}%
\providecommand \href@noop [0]{\@secondoftwo}%
\providecommand \href [0]{\begingroup \@sanitize@url \@href}%
\providecommand \@href[1]{\@@startlink{#1}\@@href}%
\providecommand \@@href[1]{\endgroup#1\@@endlink}%
\providecommand \@sanitize@url [0]{\catcode `\\12\catcode `\$12\catcode
  `\&12\catcode `\#12\catcode `\^12\catcode `\_12\catcode `\%12\relax}%
\providecommand \@@startlink[1]{}%
\providecommand \@@endlink[0]{}%
\providecommand \url  [0]{\begingroup\@sanitize@url \@url }%
\providecommand \@url [1]{\endgroup\@href {#1}{\urlprefix }}%
\providecommand \urlprefix  [0]{URL }%
\providecommand \Eprint [0]{\href }%
\providecommand \doibase [0]{http://dx.doi.org/}%
\providecommand \selectlanguage [0]{\@gobble}%
\providecommand \bibinfo  [0]{\@secondoftwo}%
\providecommand \bibfield  [0]{\@secondoftwo}%
\providecommand \translation [1]{[#1]}%
\providecommand \BibitemOpen [0]{}%
\providecommand \bibitemStop [0]{}%
\providecommand \bibitemNoStop [0]{.\EOS\space}%
\providecommand \EOS [0]{\spacefactor3000\relax}%
\providecommand \BibitemShut  [1]{\csname bibitem#1\endcsname}%
\let\auto@bib@innerbib\@empty
\bibitem [{\citenamefont {Alvarez}(1989)}]{quantum-gravity}%
  \BibitemOpen
  \bibfield  {author} {\bibinfo {author} {\bibfnamefont {E.}~\bibnamefont
  {Alvarez}},\ }\href {\doibase 10.1103/RevModPhys.61.561} {\bibfield
  {journal} {\bibinfo  {journal} {Rev. Mod. Phys.}\ }\textbf {\bibinfo {volume}
  {61}},\ \bibinfo {pages} {561} (\bibinfo {year} {1989})}\BibitemShut
  {NoStop}%
\bibitem [{\citenamefont {Aad}\ \emph {et~al.}(2015)\citenamefont {Aad} \emph
  {et~al.}}]{LHC}%
  \BibitemOpen
  \bibfield  {author} {\bibinfo {author} {\bibfnamefont {G.}~\bibnamefont
  {Aad}} \emph {et~al.} (\bibinfo {collaboration} {ATLAS, CMS}),\ }\href
  {\doibase 10.1103/PhysRevLett.114.191803} {\bibfield  {journal} {\bibinfo
  {journal} {Phys. Rev. Lett.}\ }\textbf {\bibinfo {volume} {114}},\ \bibinfo
  {pages} {191803} (\bibinfo {year} {2015})},\ \Eprint
  {http://arxiv.org/abs/1503.07589} {arXiv:1503.07589 [hep-ex]} \BibitemShut
  {NoStop}%
\bibitem [{\citenamefont {Sakharov}(1968)}]{sakharov}%
  \BibitemOpen
  \bibfield  {author} {\bibinfo {author} {\bibfnamefont {A.~D.}\ \bibnamefont
  {Sakharov}},\ }\href@noop {} {\bibfield  {journal} {\bibinfo  {journal} {Sov.
  Phys. Dokl.}\ }\textbf {\bibinfo {volume} {12}},\ \bibinfo {pages} {1040}
  (\bibinfo {year} {1968})},\ \bibinfo {note} {[,51(1967)]}\BibitemShut
  {NoStop}%
\bibitem [{\citenamefont {Adler}(1980{\natexlab{a}})}]{adler1}%
  \BibitemOpen
  \bibfield  {author} {\bibinfo {author} {\bibfnamefont {S.~L.}\ \bibnamefont
  {Adler}},\ }\href {\doibase 10.1016/0370-2693(80)90478-5} {\bibfield
  {journal} {\bibinfo  {journal} {Phys. Lett.}\ }\textbf {\bibinfo {volume}
  {B95}},\ \bibinfo {pages} {241} (\bibinfo {year} {1980}{\natexlab{a}})},\
  \bibinfo {note} {[,536(1980)]}\BibitemShut {NoStop}%
\bibitem [{\citenamefont {Adler}(1980{\natexlab{b}})}]{adler2}%
  \BibitemOpen
  \bibfield  {author} {\bibinfo {author} {\bibfnamefont {S.~L.}\ \bibnamefont
  {Adler}},\ }\bibfield  {booktitle} {\emph {\bibinfo {booktitle}
  {{Proceedings, 20th International Conference on High-Energy Physics: Madison,
  Wisconsin, July 17-23, 1980}}},\ }\href {\doibase 10.1063/1.2948651}
  {\bibfield  {journal} {\bibinfo  {journal} {AIP Conf. Proc.}\ }\textbf
  {\bibinfo {volume} {68}},\ \bibinfo {pages} {915} (\bibinfo {year}
  {1980}{\natexlab{b}})}\BibitemShut {NoStop}%
\bibitem [{\citenamefont {Zee}(1983)}]{zee}%
  \BibitemOpen
  \bibfield  {author} {\bibinfo {author} {\bibfnamefont {A.}~\bibnamefont
  {Zee}},\ }\href {\doibase 10.1016/0003-4916(83)90286-5} {\bibfield  {journal}
  {\bibinfo  {journal} {Annals Phys.}\ }\textbf {\bibinfo {volume} {151}},\
  \bibinfo {pages} {431} (\bibinfo {year} {1983})}\BibitemShut {NoStop}%
\bibitem [{\citenamefont {Jordan}(1955)}]{jordan}%
  \BibitemOpen
  \bibfield  {author} {\bibinfo {author} {\bibfnamefont {P.}~\bibnamefont
  {Jordan}},\ }\href@noop {} {\bibfield  {journal} {\bibinfo  {journal}
  {Friedch Vieweg and Sohn. Braunschweig}\ ,\ \bibinfo {pages} {207}} (\bibinfo
  {year} {1955})}\BibitemShut {NoStop}%
\bibitem [{\citenamefont {Fujii}\ and\ \citenamefont {Maeda}(2007)}]{maeda}%
  \BibitemOpen
  \bibfield  {author} {\bibinfo {author} {\bibfnamefont {Y.}~\bibnamefont
  {Fujii}}\ and\ \bibinfo {author} {\bibfnamefont {K.}~\bibnamefont {Maeda}},\
  }\href {\doibase 10.1017/CBO9780511535093} {\emph {\bibinfo {title} {{The
  scalar-tensor theory of gravitation}}}},\ Cambridge Monographs on
  Mathematical Physics\ (\bibinfo  {publisher} {Cambridge University Press},\
  \bibinfo {year} {2007})\BibitemShut {NoStop}%
\bibitem [{\citenamefont {Brans}\ and\ \citenamefont {Dicke}(1961)}]{BD}%
  \BibitemOpen
  \bibfield  {author} {\bibinfo {author} {\bibfnamefont {C.}~\bibnamefont
  {Brans}}\ and\ \bibinfo {author} {\bibfnamefont {R.~H.}\ \bibnamefont
  {Dicke}},\ }\href {\doibase 10.1103/PhysRev.124.925} {\bibfield  {journal}
  {\bibinfo  {journal} {Phys. Rev.}\ }\textbf {\bibinfo {volume} {124}},\
  \bibinfo {pages} {925} (\bibinfo {year} {1961})},\ \bibinfo {note}
  {[,142(1961)]}\BibitemShut {NoStop}%
\bibitem [{\citenamefont {Wagoner}(1970)}]{wagoner}%
  \BibitemOpen
  \bibfield  {author} {\bibinfo {author} {\bibfnamefont {R.~V.}\ \bibnamefont
  {Wagoner}},\ }\href {\doibase 10.1103/PhysRevD.1.3209} {\bibfield  {journal}
  {\bibinfo  {journal} {Phys. Rev.}\ }\textbf {\bibinfo {volume} {D1}},\
  \bibinfo {pages} {3209} (\bibinfo {year} {1970})}\BibitemShut {NoStop}%
\bibitem [{\citenamefont {Bronnikov}\ \emph {et~al.}(2016)\citenamefont
  {Bronnikov}, \citenamefont {Fabris},\ and\ \citenamefont
  {Rodrigues}}]{bronnikov}%
  \BibitemOpen
  \bibfield  {author} {\bibinfo {author} {\bibfnamefont {K.~A.}\ \bibnamefont
  {Bronnikov}}, \bibinfo {author} {\bibfnamefont {J.~C.}\ \bibnamefont
  {Fabris}}, \ and\ \bibinfo {author} {\bibfnamefont {D.~C.}\ \bibnamefont
  {Rodrigues}},\ }\bibfield  {booktitle} {\emph {\bibinfo {booktitle}
  {{Proceedings, 3rd Amazonian Symposium on Physics: Belem, Brazil, September
  28-October 2, 2015}}},\ }\href {\doibase 10.1142/S0218271816410054}
  {\bibfield  {journal} {\bibinfo  {journal} {Int. J. Mod. Phys.}\ }\textbf
  {\bibinfo {volume} {D25}},\ \bibinfo {pages} {1641005} (\bibinfo {year}
  {2016})},\ \Eprint {http://arxiv.org/abs/1603.03692} {arXiv:1603.03692
  [gr-qc]} \BibitemShut {NoStop}%
\bibitem [{\citenamefont {Boisseau}\ \emph {et~al.}(2000)\citenamefont
  {Boisseau}, \citenamefont {Esposito-Farese}, \citenamefont {Polarski},\ and\
  \citenamefont {Starobinsky}}]{boisseau}%
  \BibitemOpen
  \bibfield  {author} {\bibinfo {author} {\bibfnamefont {B.}~\bibnamefont
  {Boisseau}}, \bibinfo {author} {\bibfnamefont {G.}~\bibnamefont
  {Esposito-Farese}}, \bibinfo {author} {\bibfnamefont {D.}~\bibnamefont
  {Polarski}}, \ and\ \bibinfo {author} {\bibfnamefont {A.~A.}\ \bibnamefont
  {Starobinsky}},\ }\href {\doibase 10.1103/PhysRevLett.85.2236} {\bibfield
  {journal} {\bibinfo  {journal} {Phys. Rev. Lett.}\ }\textbf {\bibinfo
  {volume} {85}},\ \bibinfo {pages} {2236} (\bibinfo {year} {2000})},\ \Eprint
  {http://arxiv.org/abs/gr-qc/0001066} {arXiv:gr-qc/0001066 [gr-qc]}
  \BibitemShut {NoStop}%
\bibitem [{\citenamefont {Cervantes-Cota}\ and\ \citenamefont
  {Dehnen}(1995)}]{early1}%
  \BibitemOpen
  \bibfield  {author} {\bibinfo {author} {\bibfnamefont {J.~L.}\ \bibnamefont
  {Cervantes-Cota}}\ and\ \bibinfo {author} {\bibfnamefont {H.}~\bibnamefont
  {Dehnen}},\ }\href {\doibase 10.1103/PhysRevD.51.395} {\bibfield  {journal}
  {\bibinfo  {journal} {Phys. Rev.}\ }\textbf {\bibinfo {volume} {D51}},\
  \bibinfo {pages} {395} (\bibinfo {year} {1995})},\ \Eprint
  {http://arxiv.org/abs/astro-ph/9412032} {arXiv:astro-ph/9412032 [astro-ph]}
  \BibitemShut {NoStop}%
\bibitem [{\citenamefont {Dehnen}\ and\ \citenamefont
  {Frommert}(1993)}]{early2}%
  \BibitemOpen
  \bibfield  {author} {\bibinfo {author} {\bibfnamefont {H.}~\bibnamefont
  {Dehnen}}\ and\ \bibinfo {author} {\bibfnamefont {H.}~\bibnamefont
  {Frommert}},\ }\href {\doibase 10.1007/BF00671794} {\bibfield  {journal}
  {\bibinfo  {journal} {Int. J. Theor. Phys.}\ }\textbf {\bibinfo {volume}
  {32}},\ \bibinfo {pages} {1135} (\bibinfo {year} {1993})}\BibitemShut
  {NoStop}%
\bibitem [{\citenamefont {Dehnen}\ \emph {et~al.}(1990)\citenamefont {Dehnen},
  \citenamefont {Frommert},\ and\ \citenamefont {Ghaboussi}}]{early3}%
  \BibitemOpen
  \bibfield  {author} {\bibinfo {author} {\bibfnamefont {H.}~\bibnamefont
  {Dehnen}}, \bibinfo {author} {\bibfnamefont {H.}~\bibnamefont {Frommert}}, \
  and\ \bibinfo {author} {\bibfnamefont {F.}~\bibnamefont {Ghaboussi}},\ }\href
  {\doibase 10.1007/BF00672029} {\bibfield  {journal} {\bibinfo  {journal}
  {Int. J. Theor. Phys.}\ }\textbf {\bibinfo {volume} {29}},\ \bibinfo {pages}
  {537} (\bibinfo {year} {1990})}\BibitemShut {NoStop}%
\bibitem [{\citenamefont {Dehnen}\ and\ \citenamefont
  {Frommert}(1991)}]{early4}%
  \BibitemOpen
  \bibfield  {author} {\bibinfo {author} {\bibfnamefont {H.}~\bibnamefont
  {Dehnen}}\ and\ \bibinfo {author} {\bibfnamefont {H.}~\bibnamefont
  {Frommert}},\ }\href {\doibase 10.1007/BF00673991} {\bibfield  {journal}
  {\bibinfo  {journal} {Int. J. Theor. Phys.}\ }\textbf {\bibinfo {volume}
  {30}},\ \bibinfo {pages} {985} (\bibinfo {year} {1991})}\BibitemShut
  {NoStop}%
\bibitem [{\citenamefont {Azri}\ and\ \citenamefont
  {Demir}(2018)}]{azri-induced}%
  \BibitemOpen
  \bibfield  {author} {\bibinfo {author} {\bibfnamefont {H.}~\bibnamefont
  {Azri}}\ and\ \bibinfo {author} {\bibfnamefont {D.}~\bibnamefont {Demir}},\
  }\href {\doibase 10.1103/PhysRevD.97.044025} {\bibfield  {journal} {\bibinfo
  {journal} {Phys. Rev.}\ }\textbf {\bibinfo {volume} {D97}},\ \bibinfo {pages}
  {044025} (\bibinfo {year} {2018})},\ \Eprint
  {http://arxiv.org/abs/1802.00590} {arXiv:1802.00590 [gr-qc]} \BibitemShut
  {NoStop}%
\bibitem [{\citenamefont {Azri}\ and\ \citenamefont
  {Demir}(2017)}]{azri-affine}%
  \BibitemOpen
  \bibfield  {author} {\bibinfo {author} {\bibfnamefont {H.}~\bibnamefont
  {Azri}}\ and\ \bibinfo {author} {\bibfnamefont {D.}~\bibnamefont {Demir}},\
  }\href {\doibase 10.1103/PhysRevD.95.124007} {\bibfield  {journal} {\bibinfo
  {journal} {Phys. Rev.}\ }\textbf {\bibinfo {volume} {D95}},\ \bibinfo {pages}
  {124007} (\bibinfo {year} {2017})},\ \Eprint
  {http://arxiv.org/abs/1705.05822} {arXiv:1705.05822 [gr-qc]} \BibitemShut
  {NoStop}%
\bibitem [{\citenamefont {Demir}(2014{\natexlab{a}})}]{demir-eddington}%
  \BibitemOpen
  \bibfield  {author} {\bibinfo {author} {\bibfnamefont {D.~A.}\ \bibnamefont
  {Demir}},\ }\href {\doibase 10.1103/PhysRevD.90.064017} {\bibfield  {journal}
  {\bibinfo  {journal} {Phys. Rev.}\ }\textbf {\bibinfo {volume} {D90}},\
  \bibinfo {pages} {064017} (\bibinfo {year} {2014}{\natexlab{a}})},\ \Eprint
  {http://arxiv.org/abs/1409.2572} {arXiv:1409.2572 [gr-qc]} \BibitemShut
  {NoStop}%
\bibitem [{\citenamefont {Azri}(2015)}]{azri-immersed}%
  \BibitemOpen
  \bibfield  {author} {\bibinfo {author} {\bibfnamefont {H.}~\bibnamefont
  {Azri}},\ }\href {\doibase 10.1088/0264-9381/32/6/065009} {\bibfield
  {journal} {\bibinfo  {journal} {Class. Quant. Grav.}\ }\textbf {\bibinfo
  {volume} {32}},\ \bibinfo {pages} {065009} (\bibinfo {year} {2015})},\
  \Eprint {http://arxiv.org/abs/1501.06177} {arXiv:1501.06177 [gr-qc]}
  \BibitemShut {NoStop}%
\bibitem [{\citenamefont {Azri}(2016)}]{azri-separate}%
  \BibitemOpen
  \bibfield  {author} {\bibinfo {author} {\bibfnamefont {H.}~\bibnamefont
  {Azri}},\ }\href {\doibase 10.1002/andp.201500270} {\bibfield  {journal}
  {\bibinfo  {journal} {Annalen Phys.}\ }\textbf {\bibinfo {volume} {528}},\
  \bibinfo {pages} {404} (\bibinfo {year} {2016})},\ \Eprint
  {http://arxiv.org/abs/1511.06600} {arXiv:1511.06600 [gr-qc]} \BibitemShut
  {NoStop}%
\bibitem [{\citenamefont {Poplawski}(2009)}]{poplawski}%
  \BibitemOpen
  \bibfield  {author} {\bibinfo {author} {\bibfnamefont {N.~J.}\ \bibnamefont
  {Poplawski}},\ }\href {\doibase 10.1007/s10701-009-9284-y} {\bibfield
  {journal} {\bibinfo  {journal} {Found. Phys.}\ }\textbf {\bibinfo {volume}
  {39}},\ \bibinfo {pages} {307} (\bibinfo {year} {2009})},\ \Eprint
  {http://arxiv.org/abs/gr-qc/0701176} {arXiv:gr-qc/0701176 [GR-QC]}
  \BibitemShut {NoStop}%
\bibitem [{\citenamefont {Cervantes-Cota}\ and\ \citenamefont
  {Liebscher}(2016)}]{cota}%
  \BibitemOpen
  \bibfield  {author} {\bibinfo {author} {\bibfnamefont {J.~L.}\ \bibnamefont
  {Cervantes-Cota}}\ and\ \bibinfo {author} {\bibfnamefont {D.~E.}\
  \bibnamefont {Liebscher}},\ }\href {\doibase 10.1007/s10714-016-2103-9}
  {\bibfield  {journal} {\bibinfo  {journal} {Gen. Rel. Grav.}\ }\textbf
  {\bibinfo {volume} {48}},\ \bibinfo {pages} {108} (\bibinfo {year} {2016})},\
  \Eprint {http://arxiv.org/abs/1607.04250} {arXiv:1607.04250 [gr-qc]}
  \BibitemShut {NoStop}%
\bibitem [{\citenamefont {Castillo-Felisola}\ and\ \citenamefont
  {Skirzewski}(2018)}]{oscar}%
  \BibitemOpen
  \bibfield  {author} {\bibinfo {author} {\bibfnamefont {O.}~\bibnamefont
  {Castillo-Felisola}}\ and\ \bibinfo {author} {\bibfnamefont {A.}~\bibnamefont
  {Skirzewski}},\ }\href {\doibase 10.1088/1361-6382/aaa699} {\bibfield
  {journal} {\bibinfo  {journal} {Class. Quant. Grav.}\ }\textbf {\bibinfo
  {volume} {35}},\ \bibinfo {pages} {055012} (\bibinfo {year} {2018})},\
  \Eprint {http://arxiv.org/abs/1505.04634} {arXiv:1505.04634 [gr-qc]}
  \BibitemShut {NoStop}%
\bibitem [{\citenamefont {Skirzewski}\ and\ \citenamefont
  {Castillo-Felisola}(2015)}]{oscar2}%
  \BibitemOpen
  \bibfield  {author} {\bibinfo {author} {\bibfnamefont {A.}~\bibnamefont
  {Skirzewski}}\ and\ \bibinfo {author} {\bibfnamefont {O.}~\bibnamefont
  {Castillo-Felisola}},\ }\href@noop {} {\bibfield  {journal} {\bibinfo
  {journal} {Rev. Mex. Fis.}\ }\textbf {\bibinfo {volume} {61}},\ \bibinfo
  {pages} {6} (\bibinfo {year} {2015})},\ \Eprint
  {http://arxiv.org/abs/1410.6183} {arXiv:1410.6183 [gr-qc]} \BibitemShut
  {NoStop}%
\bibitem [{\citenamefont {Kijowski}\ and\ \citenamefont
  {Werpachowski}(2007)}]{kijowski1}%
  \BibitemOpen
  \bibfield  {author} {\bibinfo {author} {\bibfnamefont {J.}~\bibnamefont
  {Kijowski}}\ and\ \bibinfo {author} {\bibfnamefont {R.}~\bibnamefont
  {Werpachowski}},\ }\href {\doibase 10.1016/S0034-4877(07)80001-2} {\bibfield
  {journal} {\bibinfo  {journal} {Rept. Math. Phys.}\ }\textbf {\bibinfo
  {volume} {59}},\ \bibinfo {pages} {1} (\bibinfo {year} {2007})},\ \Eprint
  {http://arxiv.org/abs/gr-qc/0406088} {arXiv:gr-qc/0406088 [gr-qc]}
  \BibitemShut {NoStop}%
\bibitem [{\citenamefont {Kijowski}(1978)}]{kijowski2}%
  \BibitemOpen
  \bibfield  {author} {\bibinfo {author} {\bibfnamefont {J.}~\bibnamefont
  {Kijowski}},\ }\href@noop {} {\bibfield  {journal} {\bibinfo  {journal}
  {Gen.Rel.Grav.}\ }\textbf {\bibinfo {volume} {9}},\ \bibinfo {pages} {857}
  (\bibinfo {year} {1978})}\BibitemShut {NoStop}%
\bibitem [{\citenamefont {Azri}(2018)}]{azri-review}%
  \BibitemOpen
  \bibfield  {author} {\bibinfo {author} {\bibfnamefont {H.}~\bibnamefont
  {Azri}},\ }\href {\doibase 10.1142/S0218271818300069} {\bibfield  {journal}
  {\bibinfo  {journal} {Int. J. Mod. Phys.}\ }\textbf {\bibinfo {volume}
  {D27}},\ \bibinfo {pages} {1830006} (\bibinfo {year} {2018})},\ \Eprint
  {http://arxiv.org/abs/1802.01247} {arXiv:1802.01247 [gr-qc]} \BibitemShut
  {NoStop}%
\bibitem [{\citenamefont {Guth}(1981)}]{guth}%
  \BibitemOpen
  \bibfield  {author} {\bibinfo {author} {\bibfnamefont {A.~H.}\ \bibnamefont
  {Guth}},\ }\href {\doibase 10.1103/PhysRevD.23.347} {\bibfield  {journal}
  {\bibinfo  {journal} {Phys. Rev.}\ }\textbf {\bibinfo {volume} {D23}},\
  \bibinfo {pages} {347} (\bibinfo {year} {1981})}\BibitemShut {NoStop}%
\bibitem [{\citenamefont {Linde}(1982)}]{linde1}%
  \BibitemOpen
  \bibfield  {author} {\bibinfo {author} {\bibfnamefont {A.~D.}\ \bibnamefont
  {Linde}},\ }\bibfield  {booktitle} {\emph {\bibinfo {booktitle} {{Second
  Seminar on Quantum Gravity Moscow, USSR, October 13-15, 1981}}},\ }\href
  {\doibase 10.1016/0370-2693(82)91219-9} {\bibfield  {journal} {\bibinfo
  {journal} {Phys. Lett.}\ }\textbf {\bibinfo {volume} {108B}},\ \bibinfo
  {pages} {389} (\bibinfo {year} {1982})}\BibitemShut {NoStop}%
\bibitem [{\citenamefont {Albrecht}\ and\ \citenamefont
  {Steinhardt}(1982)}]{albrecht}%
  \BibitemOpen
  \bibfield  {author} {\bibinfo {author} {\bibfnamefont {A.}~\bibnamefont
  {Albrecht}}\ and\ \bibinfo {author} {\bibfnamefont {P.~J.}\ \bibnamefont
  {Steinhardt}},\ }\href {\doibase 10.1103/PhysRevLett.48.1220} {\bibfield
  {journal} {\bibinfo  {journal} {Phys. Rev. Lett.}\ }\textbf {\bibinfo
  {volume} {48}},\ \bibinfo {pages} {1220} (\bibinfo {year}
  {1982})}\BibitemShut {NoStop}%
\bibitem [{\citenamefont {Linde}(1983)}]{linde2}%
  \BibitemOpen
  \bibfield  {author} {\bibinfo {author} {\bibfnamefont {A.~D.}\ \bibnamefont
  {Linde}},\ }\href {\doibase 10.1016/0370-2693(83)90837-7} {\bibfield
  {journal} {\bibinfo  {journal} {Phys. Lett.}\ }\textbf {\bibinfo {volume}
  {129B}},\ \bibinfo {pages} {177} (\bibinfo {year} {1983})}\BibitemShut
  {NoStop}%
\bibitem [{\citenamefont {Bezrukov}\ and\ \citenamefont
  {Shaposhnikov}(2008)}]{higgs-inflation}%
  \BibitemOpen
  \bibfield  {author} {\bibinfo {author} {\bibfnamefont {F.~L.}\ \bibnamefont
  {Bezrukov}}\ and\ \bibinfo {author} {\bibfnamefont {M.}~\bibnamefont
  {Shaposhnikov}},\ }\href {\doibase 10.1016/j.physletb.2007.11.072} {\bibfield
   {journal} {\bibinfo  {journal} {Phys. Lett.}\ }\textbf {\bibinfo {volume}
  {B659}},\ \bibinfo {pages} {703} (\bibinfo {year} {2008})},\ \Eprint
  {http://arxiv.org/abs/0710.3755} {arXiv:0710.3755 [hep-th]} \BibitemShut
  {NoStop}%
\bibitem [{\citenamefont {Bauer}\ and\ \citenamefont {Demir}(2008)}]{bauer1}%
  \BibitemOpen
  \bibfield  {author} {\bibinfo {author} {\bibfnamefont {F.}~\bibnamefont
  {Bauer}}\ and\ \bibinfo {author} {\bibfnamefont {D.~A.}\ \bibnamefont
  {Demir}},\ }\href {\doibase 10.1016/j.physletb.2008.06.014} {\bibfield
  {journal} {\bibinfo  {journal} {Phys. Lett.}\ }\textbf {\bibinfo {volume}
  {B665}},\ \bibinfo {pages} {222} (\bibinfo {year} {2008})},\ \Eprint
  {http://arxiv.org/abs/0803.2664} {arXiv:0803.2664 [hep-ph]} \BibitemShut
  {NoStop}%
\bibitem [{\citenamefont {Bauer}\ and\ \citenamefont {Demir}(2011)}]{bauer2}%
  \BibitemOpen
  \bibfield  {author} {\bibinfo {author} {\bibfnamefont {F.}~\bibnamefont
  {Bauer}}\ and\ \bibinfo {author} {\bibfnamefont {D.~A.}\ \bibnamefont
  {Demir}},\ }\href {\doibase 10.1016/j.physletb.2011.03.042} {\bibfield
  {journal} {\bibinfo  {journal} {Phys. Lett.}\ }\textbf {\bibinfo {volume}
  {B698}},\ \bibinfo {pages} {425} (\bibinfo {year} {2011})},\ \Eprint
  {http://arxiv.org/abs/1012.2900} {arXiv:1012.2900 [hep-ph]} \BibitemShut
  {NoStop}%
\bibitem [{\citenamefont {Eddington}(1919)}]{eddington}%
  \BibitemOpen
  \bibfield  {author} {\bibinfo {author} {\bibfnamefont {A.~S.}\ \bibnamefont
  {Eddington}},\ }\href@noop {} {\bibfield  {journal} {\bibinfo  {journal}
  {Proc. Roy. Soc. Lond.}\ }\textbf {\bibinfo {volume} {A99}},\ \bibinfo
  {pages} {742} (\bibinfo {year} {1919})}\BibitemShut {NoStop}%
\bibitem [{\citenamefont {Schrodinger}(1950)}]{schrodinger}%
  \BibitemOpen
  \bibfield  {author} {\bibinfo {author} {\bibfnamefont {E.}~\bibnamefont
  {Schrodinger}},\ }\href@noop {} {\emph {\bibinfo {title} {{Space-Time
  Structure, Cambridge University Press}}}}\ (\bibinfo {year} {1950})\ p.\
  \bibinfo {pages} {119}\BibitemShut {NoStop}%
\bibitem [{\citenamefont {Peskin}\ and\ \citenamefont
  {Schroeder}(1995)}]{peskin}%
  \BibitemOpen
  \bibfield  {author} {\bibinfo {author} {\bibfnamefont {M.~E.}\ \bibnamefont
  {Peskin}}\ and\ \bibinfo {author} {\bibfnamefont {D.~V.}\ \bibnamefont
  {Schroeder}},\ }\href {http://www.slac.stanford.edu/~mpeskin/QFT.html} {\emph
  {\bibinfo {title} {{An Introduction to quantum field theory}}}}\ (\bibinfo
  {publisher} {Addison-Wesley},\ \bibinfo {address} {Reading, USA},\ \bibinfo
  {year} {1995})\BibitemShut {NoStop}%
\bibitem [{\citenamefont {Birrell}\ and\ \citenamefont
  {Davies}(1984)}]{davies}%
  \BibitemOpen
  \bibfield  {author} {\bibinfo {author} {\bibfnamefont {N.~D.}\ \bibnamefont
  {Birrell}}\ and\ \bibinfo {author} {\bibfnamefont {P.~C.~W.}\ \bibnamefont
  {Davies}},\ }\href {\doibase 10.1017/CBO9780511622632} {\emph {\bibinfo
  {title} {{Quantum Fields in Curved Space}}}},\ Cambridge Monographs on
  Mathematical Physics\ (\bibinfo  {publisher} {Cambridge Univ. Press},\
  \bibinfo {address} {Cambridge, UK},\ \bibinfo {year} {1984})\BibitemShut
  {NoStop}%
\bibitem [{\citenamefont {Buchbinder}\ \emph {et~al.}(1992)\citenamefont
  {Buchbinder}, \citenamefont {Odintsov},\ and\ \citenamefont
  {Shapiro}}]{buch}%
  \BibitemOpen
  \bibfield  {author} {\bibinfo {author} {\bibfnamefont {I.~L.}\ \bibnamefont
  {Buchbinder}}, \bibinfo {author} {\bibfnamefont {S.~D.}\ \bibnamefont
  {Odintsov}}, \ and\ \bibinfo {author} {\bibfnamefont {I.~L.}\ \bibnamefont
  {Shapiro}},\ }\href@noop {} {\emph {\bibinfo {title} {{Effective action in
  quantum gravity}}}}\ (\bibinfo {year} {1992})\BibitemShut {NoStop}%
\bibitem [{\citenamefont {Visser}(2002)}]{visser}%
  \BibitemOpen
  \bibfield  {author} {\bibinfo {author} {\bibfnamefont {M.}~\bibnamefont
  {Visser}},\ }\bibfield  {booktitle} {\emph {\bibinfo {booktitle} {{The
  interface of gravitational and quantum realms. Proceedings, 1st
  Inter-University Centre for Astronomy and Astrophysics Meeting, Pune, India,
  December 17-21, 2001}}},\ }\href {\doibase 10.1142/S0217732302006886}
  {\bibfield  {journal} {\bibinfo  {journal} {Mod. Phys. Lett.}\ }\textbf
  {\bibinfo {volume} {A17}},\ \bibinfo {pages} {977} (\bibinfo {year}
  {2002})},\ \Eprint {http://arxiv.org/abs/gr-qc/0204062} {arXiv:gr-qc/0204062
  [gr-qc]} \BibitemShut {NoStop}%
\bibitem [{\citenamefont {Chaichian}\ \emph {et~al.}(2018)\citenamefont
  {Chaichian}, \citenamefont {Oksanen},\ and\ \citenamefont
  {Tureanu}}]{extended-sakharov}%
  \BibitemOpen
  \bibfield  {author} {\bibinfo {author} {\bibfnamefont {M.}~\bibnamefont
  {Chaichian}}, \bibinfo {author} {\bibfnamefont {M.}~\bibnamefont {Oksanen}},
  \ and\ \bibinfo {author} {\bibfnamefont {A.}~\bibnamefont {Tureanu}},\
  }\href@noop {} {\  (\bibinfo {year} {2018})},\ \Eprint
  {http://arxiv.org/abs/1805.03148} {arXiv:1805.03148 [hep-th]} \BibitemShut
  {NoStop}%
\bibitem [{\citenamefont {Ade}\ \emph {et~al.}(2016)\citenamefont {Ade} \emph
  {et~al.}}]{planck}%
  \BibitemOpen
  \bibfield  {author} {\bibinfo {author} {\bibfnamefont {P.~A.~R.}\
  \bibnamefont {Ade}} \emph {et~al.} (\bibinfo {collaboration} {Planck}),\
  }\href {\doibase 10.1051/0004-6361/201525830} {\bibfield  {journal} {\bibinfo
   {journal} {Astron. Astrophys.}\ }\textbf {\bibinfo {volume} {594}},\
  \bibinfo {pages} {A13} (\bibinfo {year} {2016})},\ \Eprint
  {http://arxiv.org/abs/1502.01589} {arXiv:1502.01589 [astro-ph.CO]}
  \BibitemShut {NoStop}%
\bibitem [{\citenamefont {Zel'dovich}\ \emph {et~al.}(1968)\citenamefont
  {Zel'dovich}, \citenamefont {Krasinski},\ and\ \citenamefont
  {Zeldovich}}]{ccp1}%
  \BibitemOpen
  \bibfield  {author} {\bibinfo {author} {\bibfnamefont {{\relax Ya}.~B.}\
  \bibnamefont {Zel'dovich}}, \bibinfo {author} {\bibfnamefont
  {A.}~\bibnamefont {Krasinski}}, \ and\ \bibinfo {author} {\bibfnamefont
  {{\relax Ya}.~B.}\ \bibnamefont {Zeldovich}},\ }\href {\doibase
  10.1007/s10714-008-0624-6, 10.1070/PU1968v011n03ABEH003927} {\bibfield
  {journal} {\bibinfo  {journal} {Sov. Phys. Usp.}\ }\textbf {\bibinfo {volume}
  {11}},\ \bibinfo {pages} {381} (\bibinfo {year} {1968})},\ \bibinfo {note}
  {[Usp. Fiz. Nauk95,209(1968)]}\BibitemShut {NoStop}%
\bibitem [{\citenamefont {Weinberg}(1989)}]{ccp2}%
  \BibitemOpen
  \bibfield  {author} {\bibinfo {author} {\bibfnamefont {S.}~\bibnamefont
  {Weinberg}},\ }\href {\doibase 10.1103/RevModPhys.61.1} {\bibfield  {journal}
  {\bibinfo  {journal} {Rev. Mod. Phys.}\ }\textbf {\bibinfo {volume} {61}},\
  \bibinfo {pages} {1} (\bibinfo {year} {1989})},\ \bibinfo {note}
  {[,569(1988)]}\BibitemShut {NoStop}%
\bibitem [{\citenamefont {Azri}(8 04)}]{azri-thesis}%
  \BibitemOpen
  \bibfield  {author} {\bibinfo {author} {\bibfnamefont {H.}~\bibnamefont
  {Azri}},\ }\emph {\bibinfo {title} {{Cosmological Implications of Affine
  Gravity}}},\ \href@noop {} {Ph.D. thesis},\ \bibinfo  {school} {Izmir Inst.
  Tech.} (\bibinfo {year} {2018-04}),\ \Eprint
  {http://arxiv.org/abs/1805.03936} {arXiv:1805.03936 [gr-qc]} \BibitemShut
  {NoStop}%
\bibitem [{\citenamefont {Kurkov}\ and\ \citenamefont
  {Sakellariadou}(2014)}]{extended-sakharov1}%
  \BibitemOpen
  \bibfield  {author} {\bibinfo {author} {\bibfnamefont {M.~A.}\ \bibnamefont
  {Kurkov}}\ and\ \bibinfo {author} {\bibfnamefont {M.}~\bibnamefont
  {Sakellariadou}},\ }\href {\doibase 10.1088/1475-7516/2014/01/035} {\bibfield
   {journal} {\bibinfo  {journal} {JCAP}\ }\textbf {\bibinfo {volume} {1401}},\
  \bibinfo {pages} {035} (\bibinfo {year} {2014})},\ \Eprint
  {http://arxiv.org/abs/1311.6979} {arXiv:1311.6979 [hep-th]} \BibitemShut
  {NoStop}%
\bibitem [{\citenamefont {Demir}(2019)}]{demir-uv0}%
  \BibitemOpen
  \bibfield  {author} {\bibinfo {author} {\bibfnamefont {D.}~\bibnamefont
  {Demir}},\ }\href@noop {} {\  (\bibinfo {year} {2019})},\ \Eprint
  {http://arxiv.org/abs/1901.07244} {arXiv:1901.07244 [hep-ph]} \BibitemShut
  {NoStop}%
\bibitem [{\citenamefont {Demir}(2017)}]{demir-uv1}%
  \BibitemOpen
  \bibfield  {author} {\bibinfo {author} {\bibfnamefont {D.}~\bibnamefont
  {Demir}},\ }\href@noop {} {\  (\bibinfo {year} {2017})},\ \Eprint
  {http://arxiv.org/abs/1703.05733} {arXiv:1703.05733 [hep-ph]} \BibitemShut
  {NoStop}%
\bibitem [{\citenamefont {Demir}(2016)}]{demir-uv2}%
  \BibitemOpen
  \bibfield  {author} {\bibinfo {author} {\bibfnamefont {D.~A.}\ \bibnamefont
  {Demir}},\ }\href {\doibase 10.1155/2016/6727805} {\bibfield  {journal}
  {\bibinfo  {journal} {Adv. High Energy Phys.}\ }\textbf {\bibinfo {volume}
  {2016}},\ \bibinfo {pages} {6727805} (\bibinfo {year} {2016})},\ \Eprint
  {http://arxiv.org/abs/1605.00377} {arXiv:1605.00377 [hep-ph]} \BibitemShut
  {NoStop}%
\bibitem [{\citenamefont {Demir}(2015)}]{demir-uv3}%
  \BibitemOpen
  \bibfield  {author} {\bibinfo {author} {\bibfnamefont {D.~A.}\ \bibnamefont
  {Demir}},\ }\href@noop {} {\  (\bibinfo {year} {2015})},\ \Eprint
  {http://arxiv.org/abs/1510.05570} {arXiv:1510.05570 [hep-ph]} \BibitemShut
  {NoStop}%
\bibitem [{\citenamefont {Demir}(2014{\natexlab{b}})}]{demir-uv4}%
  \BibitemOpen
  \bibfield  {author} {\bibinfo {author} {\bibfnamefont {D.~A.}\ \bibnamefont
  {Demir}},\ }\href {\doibase 10.1016/j.physletb.2014.05.002} {\bibfield
  {journal} {\bibinfo  {journal} {Phys. Lett.}\ }\textbf {\bibinfo {volume}
  {B733}},\ \bibinfo {pages} {237} (\bibinfo {year} {2014}{\natexlab{b}})},\
  \Eprint {http://arxiv.org/abs/1405.0300} {arXiv:1405.0300 [hep-ph]}
  \BibitemShut {NoStop}%
\end{thebibliography}%

\end{document}